\documentclass[ %
aps,
prb,
10pt,
superscriptaddress,
amsmath, amssymb,
]{revtex4-2}
\usepackage{graphicx}	

\usepackage{setspace}
\usepackage{amsmath}

\usepackage{overpic}
\usepackage[normalem]{ulem}
\usepackage[usenames]{color}
\usepackage{makeidx}
\usepackage{ascmac}
\usepackage{listings}
\usepackage{multirow,  booktabs}%
\usepackage{algorithm}
\usepackage{xcolor}
\usepackage[noEnd=false, italicComments=false, commentColor=gray!100, rightComments=true]{algpseudocodex}
\usepackage{physics2, fixdif, derivative}
\usepackage{mathtools}
\usepackage{mdframed}

\usephysicsmodule{braket}
\algnewcommand{\Ifcycle}[1]{\State \algorithmicif\ #1 \textbf{cycle}}
\algnewcommand{\Ifoneline}[2]{\State \algorithmicif\ #1: #2}

\usepackage{siunitx}
\DeclareSIUnit{\formulaunit}{\text{f.u.}}
\DeclareSIUnit{\cell}{\text{cell}}
\DeclareSIUnit{\bohrmagneton}{$\mu\textsubscript{B}$}
\DeclareSIUnit{\angstrom}{\text{\AA}}
\DeclareSIUnit{\angstromcubic}{\text{\AA}\textsuperscript{3}}
\DeclareSIUnit{\elementarycharge}{\text{\ensuremath{e}}}

\renewcommand{\vec}[1]{\mathbf{#1}}

\def\H0{H^0}

\algnewcommand{\LeftState}[1]{\Statex \hspace{\algorithmicindent}#1}
\begin{document}

\title{A Machine Learning Model for Predicting QSGW Band Gaps Using the Partial Density of States in LDA}
\author{Shota Takano}
\altaffiliation{Mitsubishi Electric Corporation, Hyogo 661-8661, Japan}
\affiliation{Division of Materials and Manufacturing Science, Graduate School of Engineering, Osaka University, Suita, Osaka 565-0871, Japan}
\author{Takao Kotani}
\email{takaokotani@gmail.com}
\affiliation{Department of Engineering, Tottori University, Tottori 680-8552, Japan}
\affiliation{Center for Spintronics Research Network, Osaka University, Toyonaka 560-8531, Japan}
\author{Masao Obata}
\affiliation{Graduate School of Natural Science and Technology, Kanazawa University,
Kanazawa, Ishikawa, 920-1192, Japan}
\author{Hitoshi Fujii}                                                                  
\affiliation{Department of Physics, Kansai Medical University, Hirakata, Osaka 573-1010, Japan}    
\author{Kazunori Sato}
\affiliation{Division of Materials and Manufacturing Science, Graduate School of Engineering, Osaka University, Suita, Osaka 565-0871, Japan}
\affiliation{Center for Spintronics Research Network, Osaka University, Toyonaka 560-8531, Japan}
\affiliation{Spintronics Research Network Division, OTRI, Osaka University, Toyonaka, Osaka 560-8531, Japan}
\author{Harutaka Saito}
\affiliation{Division of Materials and Manufacturing Science, Graduate School of Engineering, Osaka University, Suita, Osaka 565-0871, Japan}
\author{Tatsuki Oda}
\affiliation{Graduate School of Natural Science and Technology, Kanazawa University,
Kanazawa, Ishikawa, 920-1192, Japan}
\affiliation{Center for Spintronics Research Network, Osaka University, Toyonaka 560-8531, Japan}
\email{obata@cphys.s.kanazawa-u.ac.jp}

\begin{abstract}
   An accurate calculation of the band gaps for given crystal structures is highly desirable. However, conventional first-principles calculations based on density functional theory (DFT) within the local density approximation (LDA) fail to predict band gaps accurately. To address this issue, the quasiparticle self-consistent $GW$ (QSGW) method is often used, as it is one of the most reliable theoretical approaches for predicting band gaps. However, QSGW requires significant computational resources. To overcome this limitation, we propose to combine QSGW with our new machine learning model named DOSnp. DOSnp is an improvement of DOSnet [V. Fung, Nature Comm. 12, 88 (2021)], with the addition of pooling layers. In this study, we applied QSGW to 1,516 materials from the Materials Project [\texttt{https://materialsproject.org/}] and used DOSnp to predict QSGW band gaps as a function of the partial density of states (PDOS) in LDA. Our results demonstrate that DOSnp significantly outperforms linear regression approaches with linearly independent descriptor generation [\texttt{https://github.com/Hitoshi-FUJII/LIDG}]. This model is a prototype for predicting the properties of materials based on PDOS.
\end{abstract}

\maketitle

\section{Introduction}
For material discovery, such as in solar cell applications, it is often necessary to calculate band gaps for
given crystal structures. However, such calculations are challenging because standard first-principles methods, the density functional theory (DFT) in the local density approximation (LDA), or its derivatives, cannot predict band gaps. This is because the band gaps in DFT
can not be interpreted as the band gaps of electronic excitations in the independent-particle picture \cite{Kotani2007}. DFT cannot do effective masses as well. 

The first-principles $GW$ approximation originally started in the 1980s
\cite{Strinati1980,Strinati1982,Hybertsen1986,godby1988},
is a promising alternative to achieve accurate band-gap predictions.
This method is based on Hedin's many-body 
perturbation theory \cite{Hedin1965}. Among the various $GW$ methods, the quasiparticle self-consistent $GW$ 
(QSGW) method is particularly promising due to its self-consistency within the $GW$ framework
\cite{Kotani2007,Schilfgaarde2006}. QSGW inherently incorporates the so-called $U$ effect, which is usually handled in LDA+$U$ methods, making it applicable to complex materials such as NiO
\cite{kotani_spin_2008, Obata2023}, magnetic metals \cite{Obata2023}, and $3d$/$4f$ impurities 
\cite{saitoPRB2023,suzuki2023dieke}. Furthermore, QSGW is superior to hybrid functionals such as HSE06 
\cite{heyd_hybrid_2003} when the screened Coulomb interaction is place-dependent as in grain boudaries.
In contrast, HSE06 requires an ad hoc treatment, dielectric-dependent HSE06 \cite{Ness2024}.

In this study, we use QSGW implemented in the \texttt{ecalj} package \cite{Deguchi2016, Kotani2014,ecalj}. 
One significant advantage of \texttt{ecalj} is its ability to generate eigenfunctions/eigenvalues at any $\vec{k}$ points in the Brillouin zone (BZ) without relying on interpolation techniques such as Wannier interpolation
\cite{marzari_maximally_2012}. Furthermore, spin fluctuations and other effects can be incorporated on top of the QSGW results \cite{Okumura2019,Okumura2021}.

Despite such advantages in QSGW, QSGW requires significant computational resources. To address this challenge, we propose two strategies. 
The first strategy involves accelerating computations using GPUs \cite{obata2025gpu}. Recently, Ness et al. performed QSGW calculations on the InAs/Al interface (containing 138 atoms in a supercell) \cite{Ness2024}. 
The second strategy is to use machine learning. By training a model with band gaps calculated in QSGW, 
we can promptly predict QSGW band gaps for given crystal structures. In this article, we focus on the second strategy.

In this approach, we first apply QSGW to many materials to generate a dataset. To reduce computational
effort, we limit our calculations to crystal structures with a small number of atoms per primitive cell. In this study, we calculate 1,516 crystal structures obtained from the Materials Project (MP) \cite{jain2013}. We use our automated
QSGW computational system, \verb#ecalj_auto#, implemented in the package \verb #ecalj #, as explained in Appendix
\ref{app:ecalj_auto}. To ensure efficient operation of \verb#ecalj_auto#, the QSGW codes in \verb#ecalj# have been 
optimized, particularly for effective memory management. Using this system, we generate a QSGW dataset, train a
machine learning model on it, and then use the model to predict band gaps for given crystal structures.

There are possible choices for the machine learning model. 
In this study, we developed a new model DOSnp that takes the
partial density of states (PDOS) as input. In our case, we predict the QSGW band gaps using only PDOS in LDA, 
to avoid large computational efforts. In Ref.~\cite{dosnet}, V. Fung et al. developed DOSnet, a model
trained to predict molecular absorption energies from PDOS. Our model DOSnp, is a modification of DOSnet. 
The main difference lies in the pooling layer for feature vectors of atoms, as we will explain later.
In contrast to models that use varieties of material descriptors as inputs
\cite{LIDG, Kanda31122019,hayashi_prediction_nodate},
DOSnp is advantageous in simplicity because we only use PDOS as input. 
PDOS can contain features of crystal structures.

Let us explain why we do not utilize another machine learning model for crystal structures, 
the crystal graph convolutional neural network (CGCNN) \cite{xie_crystal_2018}, proposed by Xie and Grossman. 
CGCNN requires the crystal graph (CG) \cite{oganov_periodic-graph_2010} as input. 
This idea of using CG as input is more attractive from a point of view of simplicity than the methods with feature vectors \cite{seko2015} of Seko et al. However, we are concerned that representing
crystal structures by CG may lose important information of crystal structures. 
Note that CGCNN is a procedure for performing machine learing via the spectrums at the atomic nodes as intermediate quantities,
where the spectrums are generated from CG. Pay attention to the fact that the graph spectrum of CG is the density of states (DOS) of the eigenvalues for the adjacency matrix of CG. This is nothing but the DOS with an orbital per atom, where we set all the tight-binding parameters to be unity. Repeating convolutions in CGCNN resembles the recursion method in electronic structure calculations. Even if we can includes informations of atomic species during the convolution process, we guess that it just has the effect modifying the tight-binding parameters.
We may improve CGCNN by incorporating multichannel spectrums at atomic nodes with geometrical features, such as atomic distances and three-body angles. 
However, we guess that such improvements will not provide better spectra distributions at atomic nodes than those obtained by PDOS in the first-principles calculations. 
In other words, PDOS can be the most meaningful set of spectra distributed at atomic nodes for characterizing the Kohn-Sham Hamiltonian in DFT. This is why PDOS has been widely discussed to explain physical properties.

In Sec.\ref{sec:method}, we explain our method for generating datasets and assessing their quality. 
We then describe our machine learning model, DOSnp, as a modification of DOSnet. 
In Sec.\ref{sec:result}, we present our results demonstrating the superiority of DOSnp compared to a linear regression (LR) approach as a reference. Since the LR approach is used in combination with the linearly independent descriptor generation (LIDG) method \cite{Kanda31122019} developed by Fujii \cite{LIDG}, we refer to this approach as LRLIDG in the following discussion.

\section{Method}
\label{sec:method}
Our purpose is to demonstrate the quality of DOSnp. This is clarified by comparing it with a linear regression approach, LRLIDG \cite{LIDG,Kanda31122019,hayashi_prediction_nodate}, as a reference.

To achieve this, we first generate a dataset as follows. Since it is not easy to systematically collect all experimental band gaps, we instead used QSGW to calculate the band gaps. This approach assumes that QSGW provides a reasonable theoretical approximation to experimental band gaps, as discussed below.

\subsection{Dataset $\Omega$ generated in QSGW}
Since DOSnp requires PDOS in LDA as input, our first step is to generate a dataset $\Omega$:
\begin{eqnarray}
\Omega &=& \{(\text{PDOS in LDA}, E_\text{g}^\text{LDA}, E_\text{g}^\text{QSGW})_i | i = 1, \dots, N \},
\end{eqnarray}
where $N = 1,516$ in this study, as explained later. Here, $E_\text{g}^\text{QSGW}$ represents the band gaps calculated in QSGW. $E_\text{g}^\text{LDA}$ represents those calculated in LDA. Although PDOS in LDA implicitly contains information about $E_\text{g}^\text{LDA}$, we explicitly include $E_\text{g}^\text{LDA}$ in 
$\Omega$ for convenience. For PDOS, we use 16 channels ($s$, $p$, $d$, $f$) per atom in the primitive cell. 
The exchange-correlation potential in LDA is based on the VWN functional \cite{VWN}.

Since we use $E_\text{g}^\text{QSGW}$ instead of experimental band gaps, our primary goal is to demonstrate
that our machine learning model provides a computational shortcut to QSGW. For the linear regression with LIDG (LRLIDG), we generate a dataset $\Omega'$:
\begin{eqnarray}
\Omega' &=& \{(\text{Descriptors}, E_\text{g}^\text{QSGW})_i | i = 1, \dots, N \},
\end{eqnarray}
where descriptors are generated from primitive descriptors using the LIDG method \cite{LIDG}. Details of LRLIDG are provided in the Appendix \ref{app:LRLIDG}.

For simplicity, we generate $\Omega$ and $\Omega'$ without considering spin-orbit coupling (SOC) for the crystal structures obtained from the Materials Project (MP). 
Previous studies have shown that QSGW performs well in calculating band gaps for a wide range of materials. For example, Table II in Ref.~\cite{Deguchi2016} provides a systematic survey of QSGW, demonstrating that its band gaps are more reliable than those obtained using LDA. 
Although QSGW80 (20\% LDA mixing) was recommended in Ref.~\cite{Deguchi2016} to avoid a
slight overestimation of band gaps, we use QSGW without LDA mixing for simplicity throughout this study.

\begin{figure}[htpb]
  \centering
  \includegraphics[width=16cm]{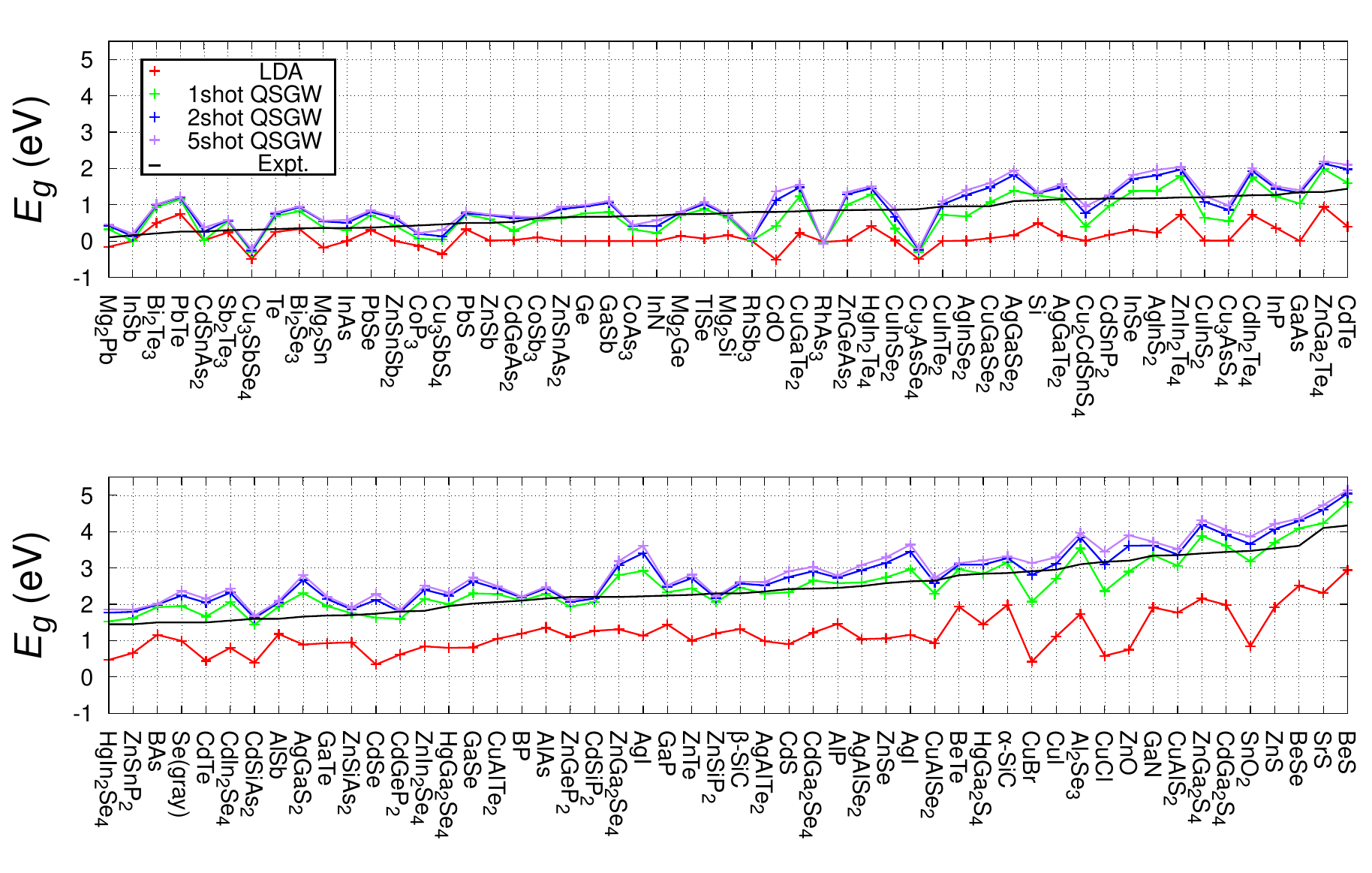}
  \caption{
    Quality check of band gaps calculated in QSGW for crystal structures in the Materials Project (MP).
    Band gaps calculated using our automated QSGW system are shown. Crystal structures are taken from MP, 
    where each material is identified by its MP ID, space group, and chemical composition.
  }
  \label{fig:bandgap0}
\end{figure}
\begin{table}[htpb]
  \centering
  \begin{tabular}{c c c c c c} \hline
     MPid  & composition & spce group & LDA & QSGW & exper. \\ \hline
     804   & GaN   & ${\rm P6_3mc}$ & 0.175 & 0.202 & 0.22\cite{GaN} \\ 
     2133  & ZnO   & ${\rm P6_3mc}$ & 0.144 & 0.250 & 0.24\cite{ZnO} \\ 
     12779 & CdTe  & ${\rm P6_3mc}$  & 0.043 & 0.114 & 0.11\cite{CdTe} \\
     560588& ZnS   & ${\rm P6_3mc}$ & 0.163 & 0.210 & 0.28\cite{ZnS} \\
     2201  & PbSe  & ${\rm Fm\overline{3}m}$ & 0.120 & 0.129 & 0.11\cite{PbSe} \\
     20351 & InP   & ${\rm F\overline{4}3m}$ & 0.032 & 0.093 & 0.073\cite{kittel} \\ 
     2534  & GaAs  & ${\rm F\overline{4}3m}$ & 0.008 & 0.069 & 0.066\cite{kittel} \\ 
     361   & ${\rm Cu_2O}$ & ${\rm Pn\overline{3}m}$ & 0.908 & 0.973 & 0.99\cite{kittel} \\ \hline
  \end{tabular}
  \caption{
   Quality check of electron effective mass $m^*$ (in units of electron mass) calculated in 
   QSGW with crystal structures in MP.
  In \texttt{ecalj}, we can calculate effective mass because 
  we can make band plot as any ${\bf k}$ point without interpolation techniques such as Wannier90.
  \cite{marzari_maximally_2012}. 
  Note that the lattice constant in MP can be different from the experimental value. 
  (For example, MP gives 5.75\AA\ for GaAs, while the experimental value is 5.65\AA).
  This can cause no band gap in LDA (VWN). 
  In all cases, the conduction band minimum (CBM) occurs at the $\Gamma$ point.
  Effective mass is evaluated with $E= \frac{\hbar^2 k^2}{2m^*}$+CBM where $k$ is measured at $E$=0.05eV+CBM in the QSGW band plot.}
  \label{tab:mass_expt}
\end{table}

With our automation scripts \verb#ecalj_auto# 
implemented in \verb#ecalj# package \cite{ecalj}, we can generate datasets
where the $k$ mesh for the self-energy parts is chosen to be 
at the level of computational quality $4\times4\times4$ for Si.
The convergence of the $k$ mesh was examined in Ref.\cite{Deguchi2016}.
In our experiences as shown in Ref.\cite{Deguchi2016} and Ref.\cite{Kotani2014},
although the mesh $4\times4\times4$ is relatively low,
it is sufficient to discuss band gaps with an accuracy of a few tenths of eV for semiconductors. 
In the case of Si, the static version of self energy (=non-local exchange-correlation potential in QSGW) is interpolated in the Brillouin zone (BZ) on a high-resolution mesh of $8\times8\times8$.
For simplicity, band gaps are evaluated on the mesh points at $8\times8\times8$ for Si.
Other materials are evaluated with the same level of $k$ mesh as well.

\subsubsection{Assesment of QSGW:}
To confirm the quality of QSGW calculation in \verb#ecalj_auto# when combined with
the crystal structures in MP, we performed the following two checks.
\begin{itemize}
  \item[(1)] Calculated band gaps are shown in Fig.~\ref{fig:bandgap0}. Experimental band gaps (black lines) are arranged in ascending order. Corresponding LDA and QSGW results are also plotted. QSGW shows a slight 
  overestimation of band gaps, consistent with Ref.~\cite{Deguchi2016}. We observe that the QSGW iteration 
  cycle gradually changes band gaps in some cases. For computational efficiency, we apply the second iteration 
  of QSGW to materials in the dataset generation. Since spin-orbit coupling (SOC) is not included, errors are 
  observed for heavy elements such as Pb. Additionally, discrepancies are noted for materials like Cu$_3$SbSe$_4$, 
  Cu$_3$SbS$_4$, RhSb$_3$, and RhAs$_3$, possibly due to issues in the crystal structures from MP or intrinsic 
  limitations of QSGW. However, these discrepancies have minimal impact on our main results, which demonstrate 
  the superiority of our machine learning model over LRLIDG.
  \item[(2)] Effective electron masses calculated in QSGW are shown in Table~\ref{tab:mass_expt}. In some cases, 
  LDA underestimates effective masses, while QSGW provides more accurate results. The crystal structures are 
  identified by their MP IDs, compositions, and space groups.
\end{itemize}

Results in Fig.~\ref{fig:bandgap0} and Table~\ref{tab:mass_expt} are consistent with those in Ref.~\cite{Deguchi2016}.

\subsubsection{Generating a dataset in QSGW:}  
We generated the dataset $\Omega$ using our automated scripts, \verb#ecalj_auto#, for crystal structures obtained 
from MP. The following conditions were applied to select the crystal structures:
\begin{itemize}
  \item Paramagnetic materials.
  \item Up to eight atoms per unit cell.
  \item No lanthanide, actinide, or noble gas elements included.
  \item Finite band gaps in PBE (Perdew-Burke-Ernzerhof) \cite{Perdew1996}.
\end{itemize}

Using these conditions, we queried MP and obtained 1,547 crystal structures. The query was performed using the 
API in pymatgen \cite{Ong2012b,Ong_2015}. Structures with no band gaps in LDA were excluded, resulting in a 
final dataset $\Omega$ containing $N = 1,516$ entries.

\begin{figure}[htbp]
   \centering
   \begin{overpic}[width=0.45\columnwidth,clip]{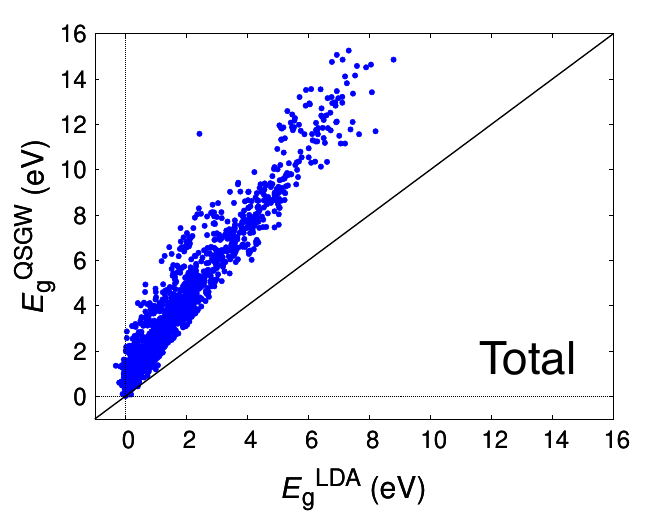}
      \put(18,67){\textbf{(a)}}
   \end{overpic}
   \hspace{0.5cm}
   \begin{overpic}[width=0.45\columnwidth,clip]{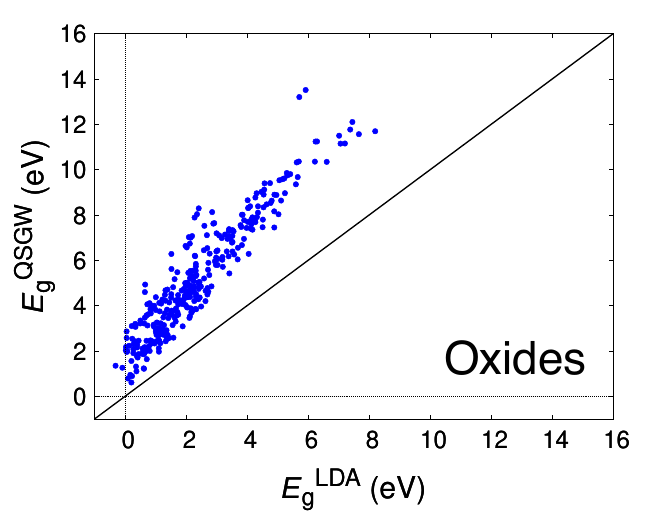}
      \put(18,67){\textbf{(b)}}
   \end{overpic} \\
   \begin{overpic}[width=0.45\columnwidth,clip]{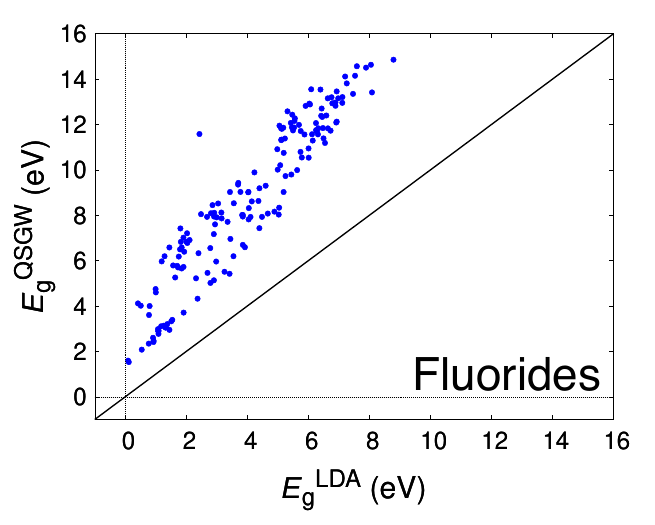}
      \put(18,67){\textbf{(c)}}
   \end{overpic}
   \hspace{0.5cm}
   \begin{overpic}[width=0.45\columnwidth,clip]{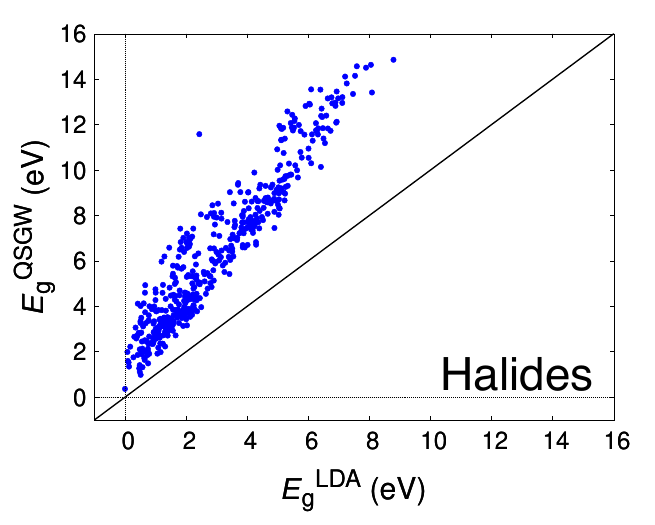}
      \put(18,67){\textbf{(d)}}
   \end{overpic}
   \caption{
      Band gap plot: LDA vs QSGW. Panel (a) shows all 1,516 data points. Panels (b), (c), and (d) show subsets 
      for oxides, fluorides, and halides, respectively. Halides include compounds containing F, Cl, Br, or I. 
      Minor differences are observed among the panels. A molecular F$_2$ appears as an outlier at 
      (LDA $\sim$ 2 eV, QSGW $\sim$ 12 eV).
   }
   \label{fig:gapsummary}
\end{figure}

In Fig.~\ref{fig:gapsummary}, we plot $E_{\rm g}^{\rm QSGW}$ vs $E_{\rm g}^{\rm LDA}$ for the dataset $\Omega$, 
classifying the data into oxides, fluorides, and halides. QSGW band gaps are significantly larger than those in LDA. 
However, no meaningful differences are observed among panels (b), (c), and (d). The scattering of data points is 
relatively large. For instance, $E_{\rm g}^{\rm QSGW}$ ranges from 3 to 8 eV when $E_{\rm g}^{\rm LDA}$ is 2 eV.

The dataset is available in the supplementary material \cite{dosnpsupp}.

\subsection{Machine learning model, DOSnp}
\begin{figure}[htpb]
   \centering
   \includegraphics[width=16cm]{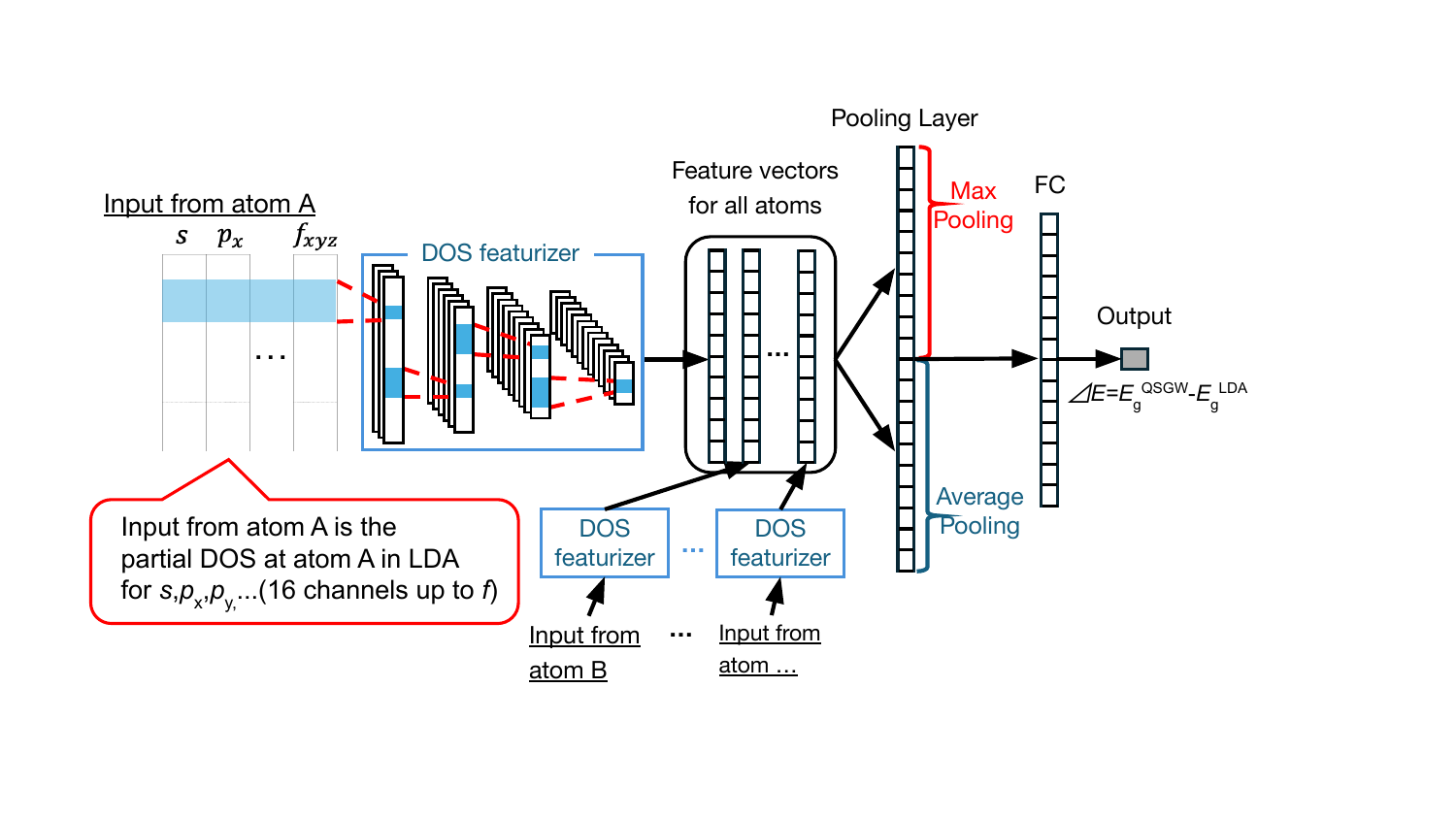}
   \caption{Schematic view of our DOSnp, as an improvement of DOSnet\cite{dosnet} to obtain 
   the difference of band gap in QSGW $E_{\rm g}^{\rm QSGW}$ and the band gap in $E_{\rm g}^{\rm LDA}$.
   Inputs are the partial density of states at the atoms in the primitive cell 
   (We take sixteen channels. Since we take PDOS up to $f$, we have $16=1+3+5+7$ channels.)
   The pooling layer, missing in DOSnet, is a key to making the output independent 
   of ordering of atoms in the primitive cell, and satisfying size scaling.
   Pooling is applied to the same rows from different atoms. Thus the size
   of pooling layer is (the size of the feature vector per atom)$\times 2$ .}
   \label{fig:DOSnp}
 \end{figure}
 Our machine learning model, DOSnp, is illustrated in Fig.~\ref{fig:DOSnp}
to obtain the difference of band gap in QSGW $E_{\rm g}^{\rm QSGW}$ and in $E_{\rm g}^{\rm LDA}$. 
The input to the model is the PDOS 
 in LDA for atoms in the primitive cell. We use 16 channels ($s$, $p$, $d$, $f$) of PDOS per atom, covering an 
 energy range of [$E_\text{F}$-1.5 Ry, $E_\text{F}$+1.0 Ry] with 1,000 divisions, where $E_\text{F}$ is the Fermi energy. 
 The architecture of our DOS featurizer is detailed in Appendix \ref{app:dosnp}.
 
 The DOS featurizer outputs a one-dimensional vector for each atom, with a size of 192. 
 We then apply max pooling and average pooling to the data (192 $\times N_a$), where $N_a$ is the number of atoms in the primitive cell. 
 This results in a one-dimensional vector of size 384, which characterizes the crystal structure. Tests show that combining max and average pooling provides better performance than using either pooling method alone.
 
 The resulting vector is passed to a fully connected (FC) perceptron layer with 100 nodes, using ReLU as the 
 activation function. A final linear transformation produces a scalar output, representing the difference 
 $E_{\rm g}^{\rm QSGW} - E_{\rm g}^{\rm LDA}$. Adding this difference to $E_{\rm g}^{\rm LDA}$ yields 
 $E_{\rm g}^{\rm QSGW}$.

DOSnp is a modification of DOSnet, with the key addition of a pooling layer. This pooling layer ensures the model is independent of the ordering of atoms in the primitive cell and satisfies size consistency, meaning that results remain unchanged when the unit cell is multiplied. As detailed in Appendix \ref{app:dosnp}, we use a relatively small model for the DOS featurizer and the subsequent FC layer compared to the original DOSnet. Although we have not yet systematically optimized the hyperparameters, the current model performs well, as shown in Sec.~\ref{sec:result}.

\section{Results}
\label{sec:result}
\begin{figure}[H]
   \centering     
   \begin{overpic}[width=0.45\columnwidth,clip]{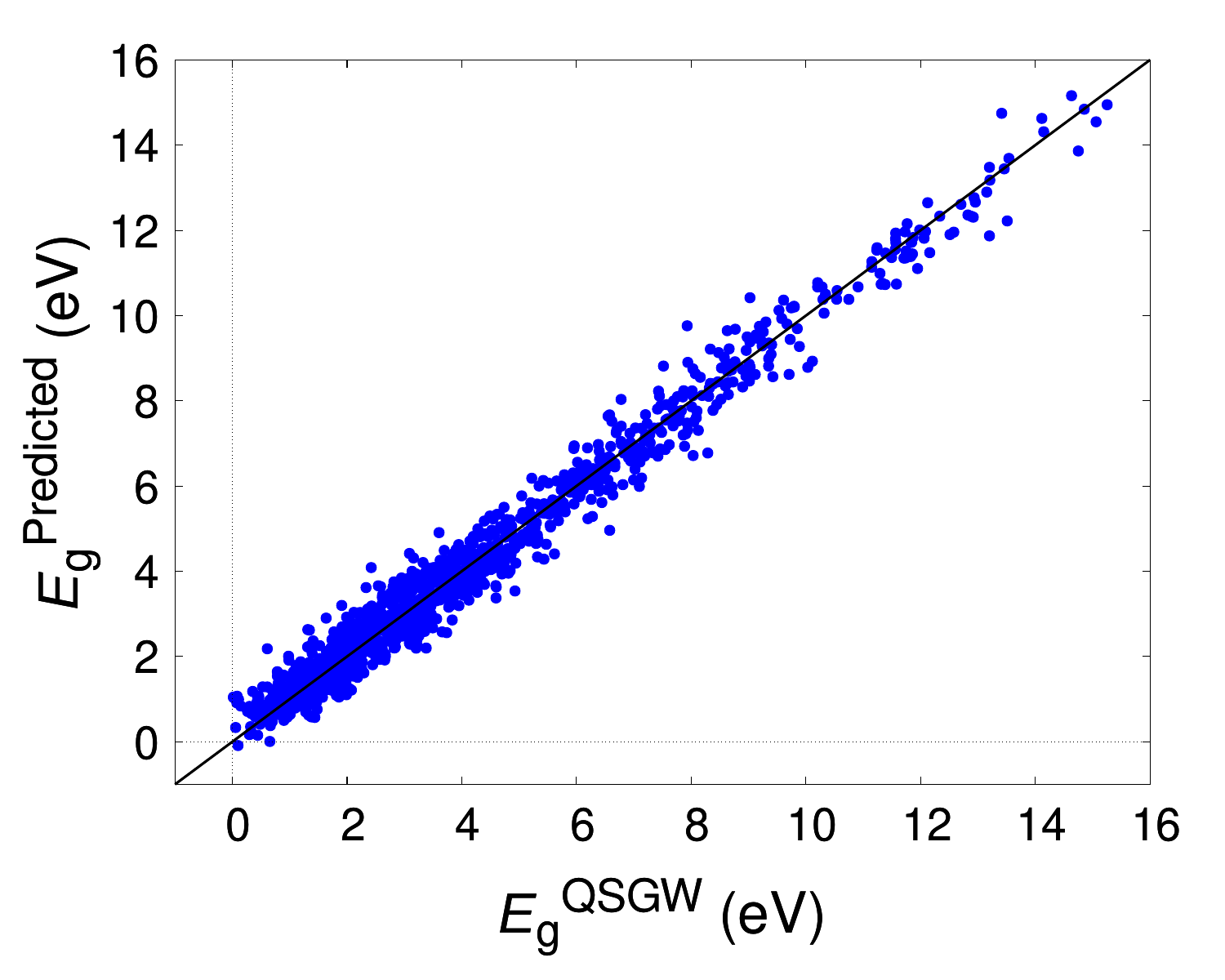}
    \put(18,68){\textbf{(a)LRLIDG training}}
 \end{overpic}
 \hspace{0.5cm}
 \begin{overpic}[width=0.45\columnwidth,clip]{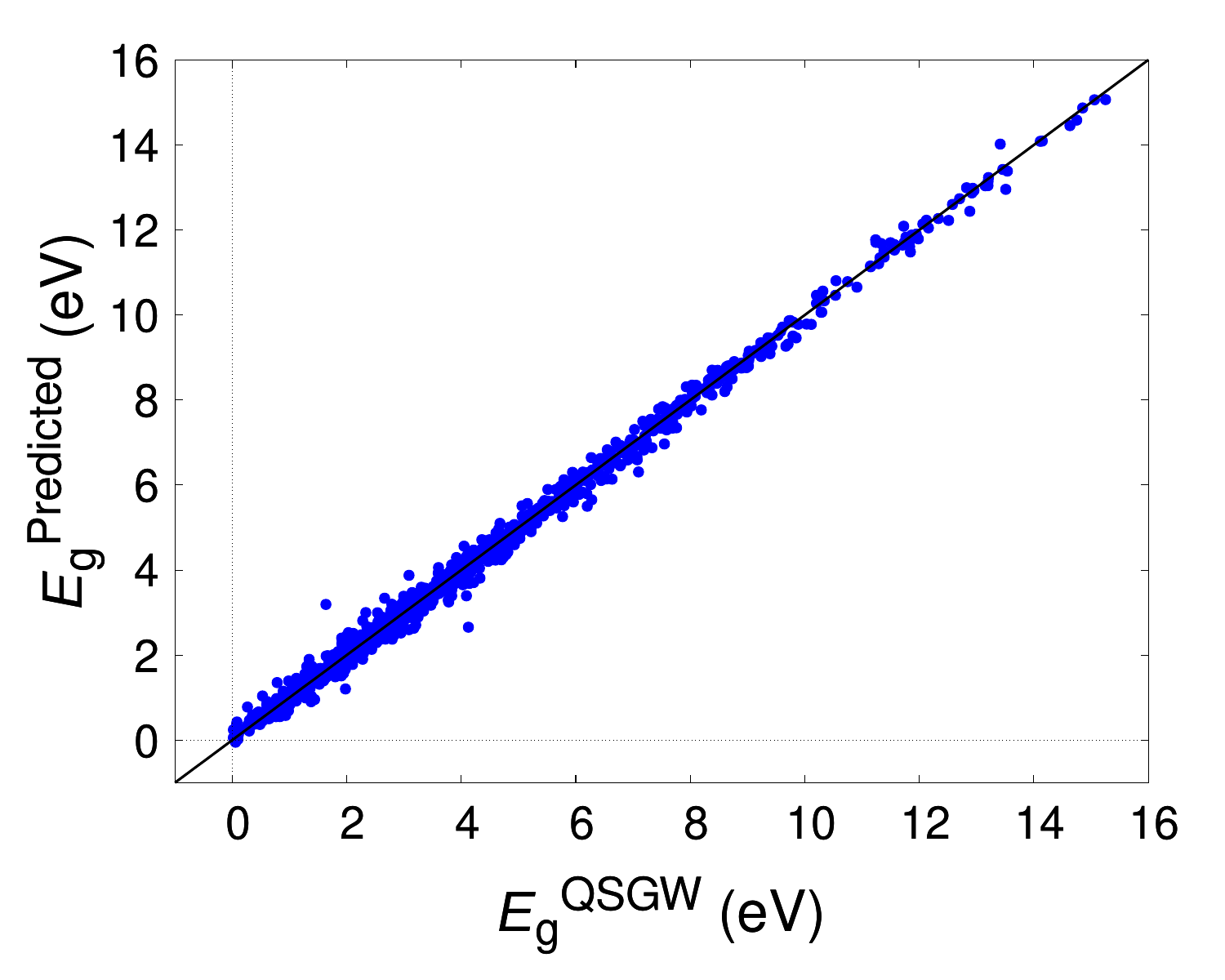}
    \put(18,68){\textbf{(A) DOSnp trainig}}
 \end{overpic}
   \begin{overpic}[width=0.45\columnwidth,clip]{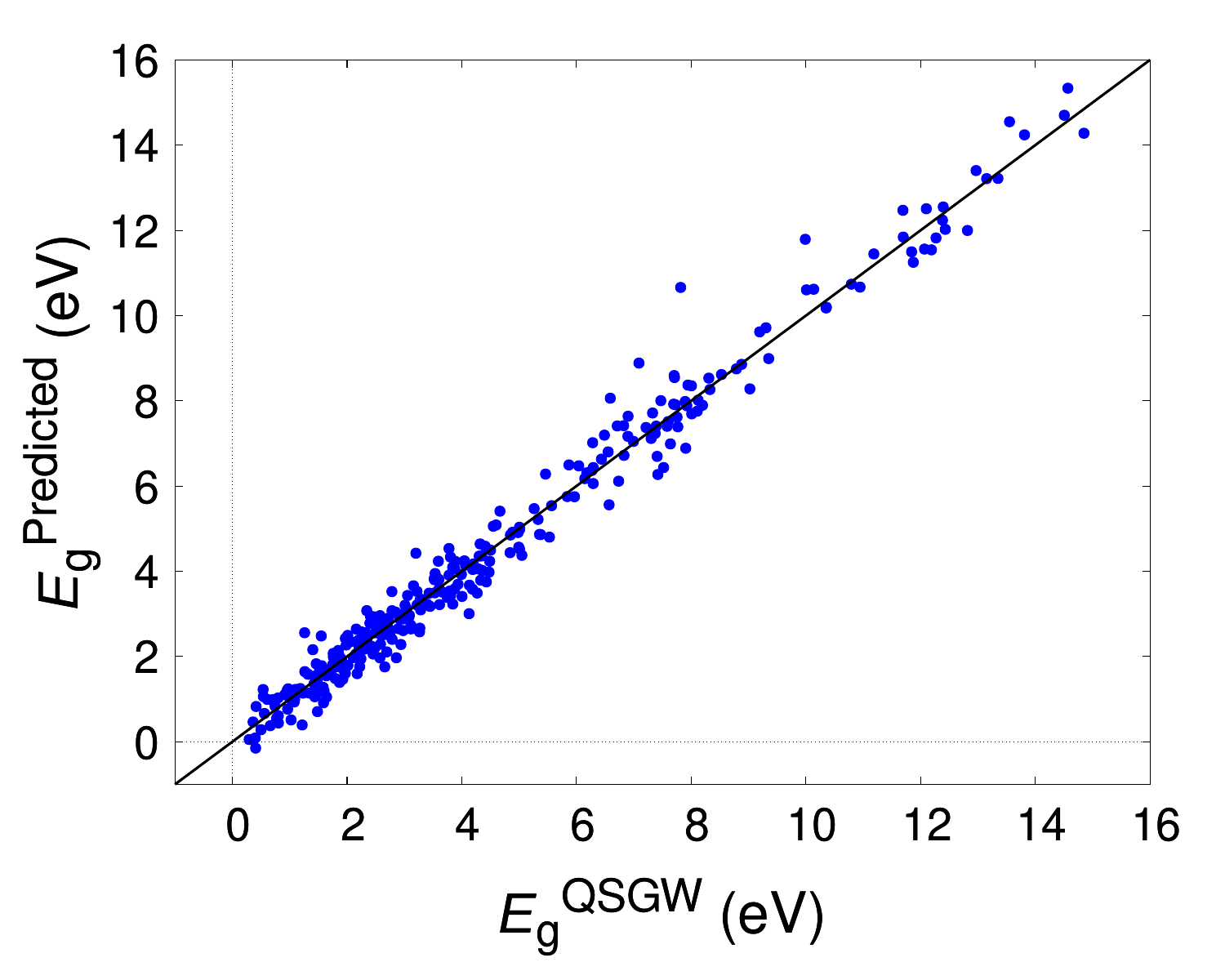}
      \put(18,68){\textbf{(b)LRLIDG test}}
   \end{overpic}
   \hspace{0.5cm}
   \begin{overpic}[width=0.45\columnwidth,clip]{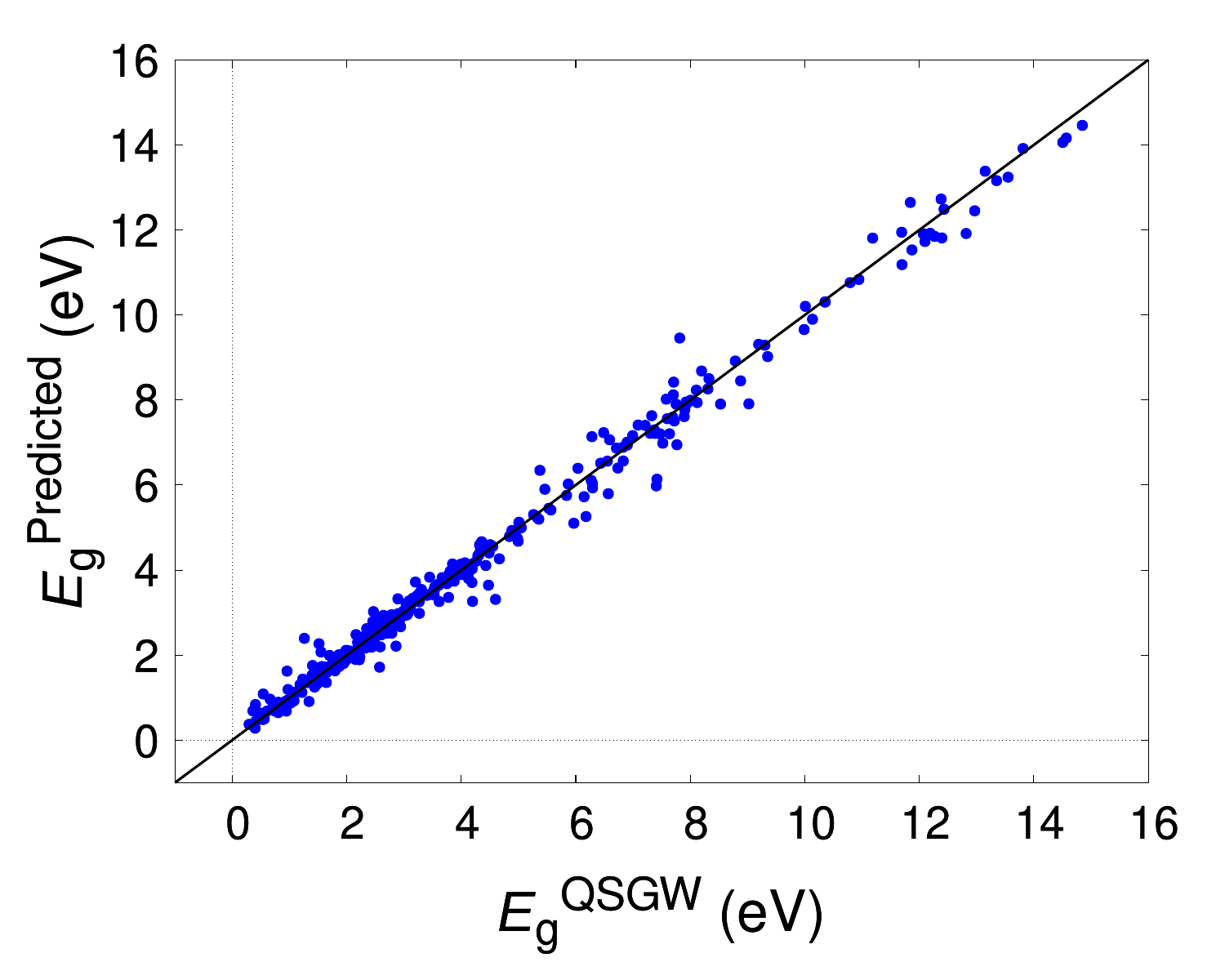}
      \put(18,68){\textbf{(B)DOSnp test}}
   \end{overpic} \\
   \begin{overpic}[width=0.45\columnwidth,clip]{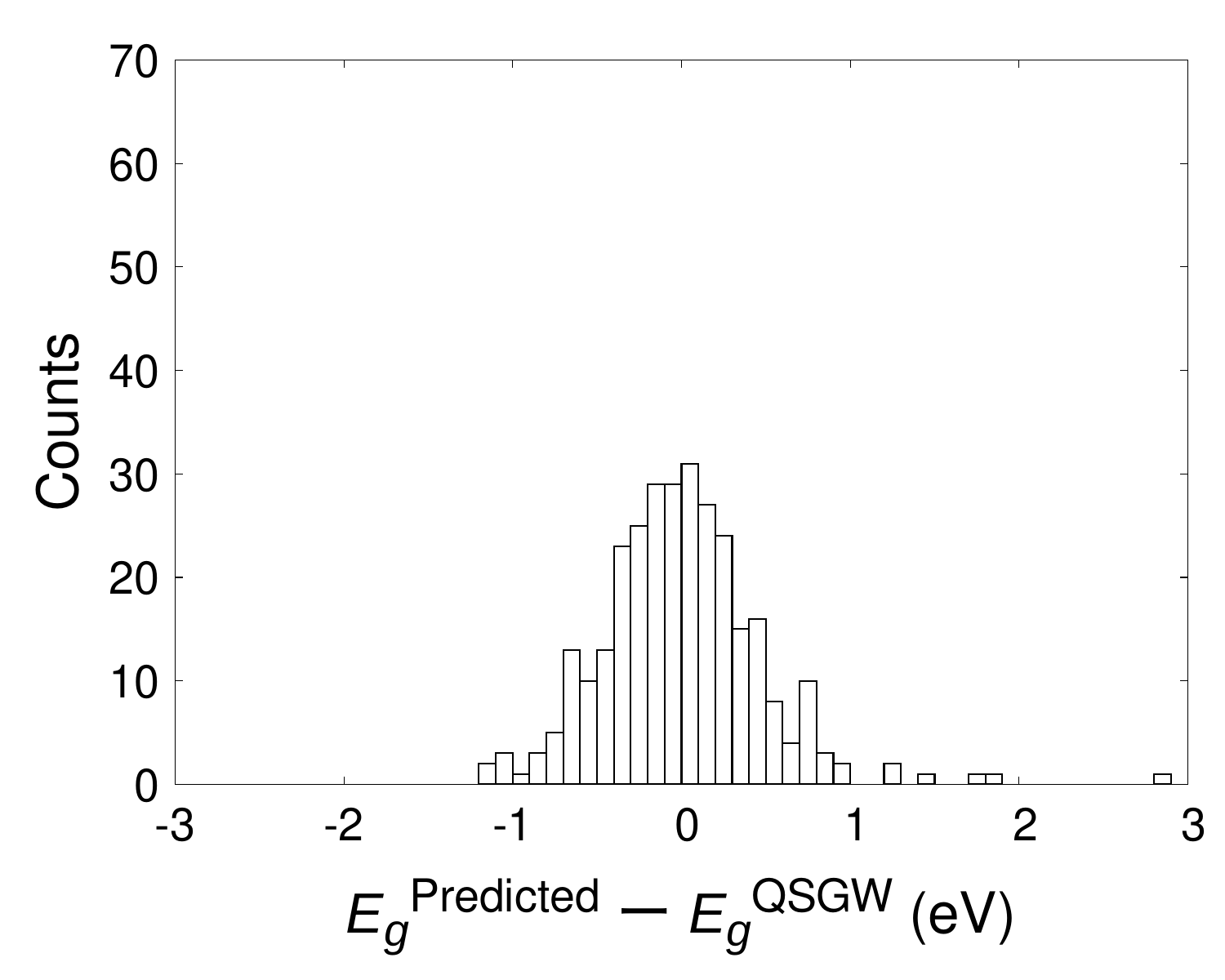}
      \put(18,68){\textbf{(c)LRLIDG test}}
   \end{overpic}
   \hspace{0.5cm}
   \begin{overpic}[width=0.45\columnwidth,clip]{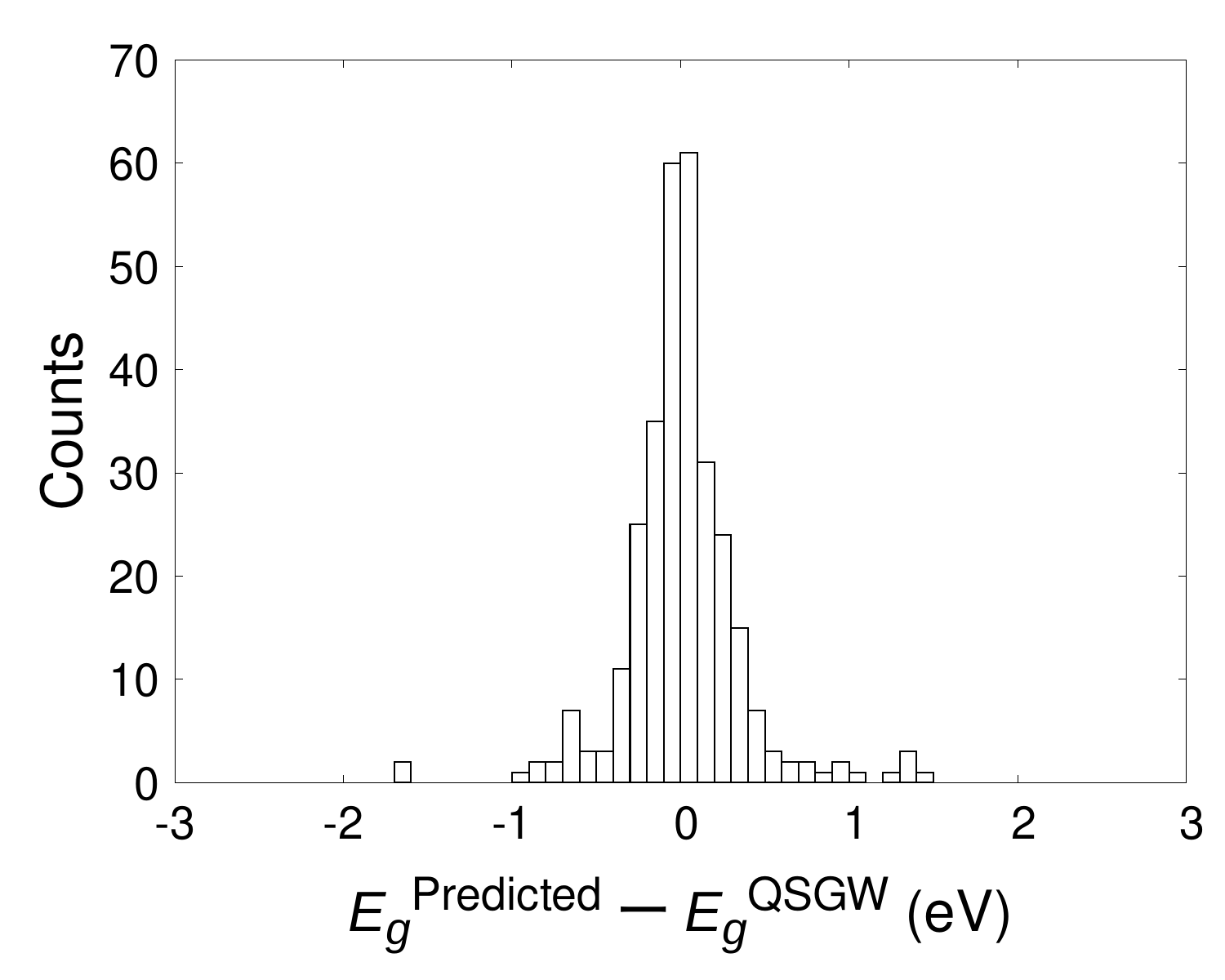}
      \put(18,68){\textbf{(C)DOSnp test}}
   \end{overpic}
   \caption{
   LRLIDG vs DOSnp for learning the band gap. The DOSnp model is explained in Sec.~\ref{sec:method}, while 
   LRLIDG is detailed in Appendix \ref{app:LRLIDG}. Training data are shown in panels (a) and (A), where the number of points is $1,516 \times 0.8$. Test results are shown in panels (b) and (B), with $1,516 \times 0.2$ 
   points. Panels (c) and (C) plot the same data as (b) and (B), respectively. DOSnp outperforms LRLIDG in the test results, as shown in (c) and (C).
}
   \label{fig:mainresults}
\end{figure}
\begin{figure}[htpb]
   \centering
   \includegraphics[width=16cm]{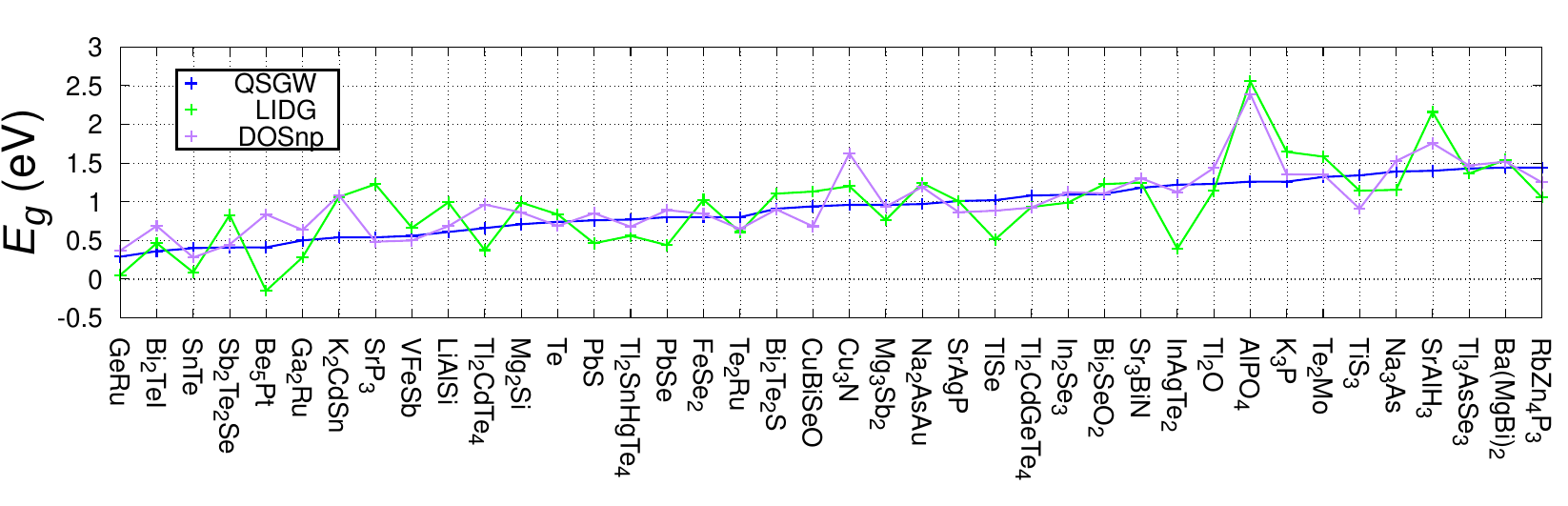}
   \caption{
      Panels (b) and (B) from Fig.~\ref{fig:mainresults} are zoomed in at low energies, with chemical compositions 
      labeled. DOSnp shows much better agreement with QSGW than LRLIDG, reflecting the differences observed 
      in panels (c) and (C) of Fig.~\ref{fig:mainresults}.
   }
   \label{fig:zoomb}
 \end{figure}

 In Fig.~\ref{fig:mainresults}, we present our main results, where the dataset of 1,516 entries is randomly divided into a training set (1,210 entries, $1512 \times 0.8$) and a test set (302 entries, $1512 \times 0.2$). Panels (a) 
 and (A) show the training data for LRLIDG and DOSnp, respectively, while panels (b) and (B) show the test results. 
 The predicted band gap, $E_{\rm g}^{\rm Predicted}$, is compared to the QSGW band gap, $E_{\rm g}^{\rm QSGW}$. 
 
 In the training results (a) and (A), $E_{\rm g}^{\rm Predicted}$ does not perfectly match $E_{\rm g}^{\rm QSGW}$ due to the limited number of learning parameters. However, the test results (b) and (B) reveal significant 
 differences in performance between LRLIDG and DOSnp. In (b), the data points for LRLIDG are highly scattered, 
 especially at low energies, and even include negative $E_{\rm g}^{\rm Predicted}$ values. In contrast, DOSnp 
 in (B) shows much less scattering and no negative predictions. 
  The error histograms in panels (c) and (C) further highlight the differences in performance. DOSnp exhibits a 
 narrower error distribution compared to LRLIDG, indicating superior accuracy.
  
In Fig.~\ref{fig:zoomb}, we zoom in on panels (b) and (B) from Fig.~\ref{fig:mainresults} at low energies, with 
chemical compositions labeled. Corresponding to panels (c) and (C) in Fig.~\ref{fig:mainresults}, DOSnp shows 
better agreement with QSGW compared to LRLIDG. However, for AlPO$_4$ (mp-4051), both models exhibit 
similar errors. This suggests that AlPO$_4$ has unique features not captured by either DOSnp or LRLIDG. AlPO$_4$ is known for its loosely packed structure, which may contribute to the observed discrepancies. This insight could provide a hint for improving DOSnp.

In addition to band gap prediction, we have extended DOSnp to predict the effective electron mass. The DOSnp model for band gaps was modified to output three values corresponding to the eigenvalues of the mass matrix. 
For simplicity, we selected a dataset containing 534 crystal structures from the original 1,516 entries, based on 
the following two conditions:
1. The conduction band minimum is at the $\Gamma$ point without degeneracy.
2. Effective masses range from 0.05 to 1.0.

\begin{figure}[H]
  \centering
  \begin{overpic}[width=0.45\columnwidth,clip]{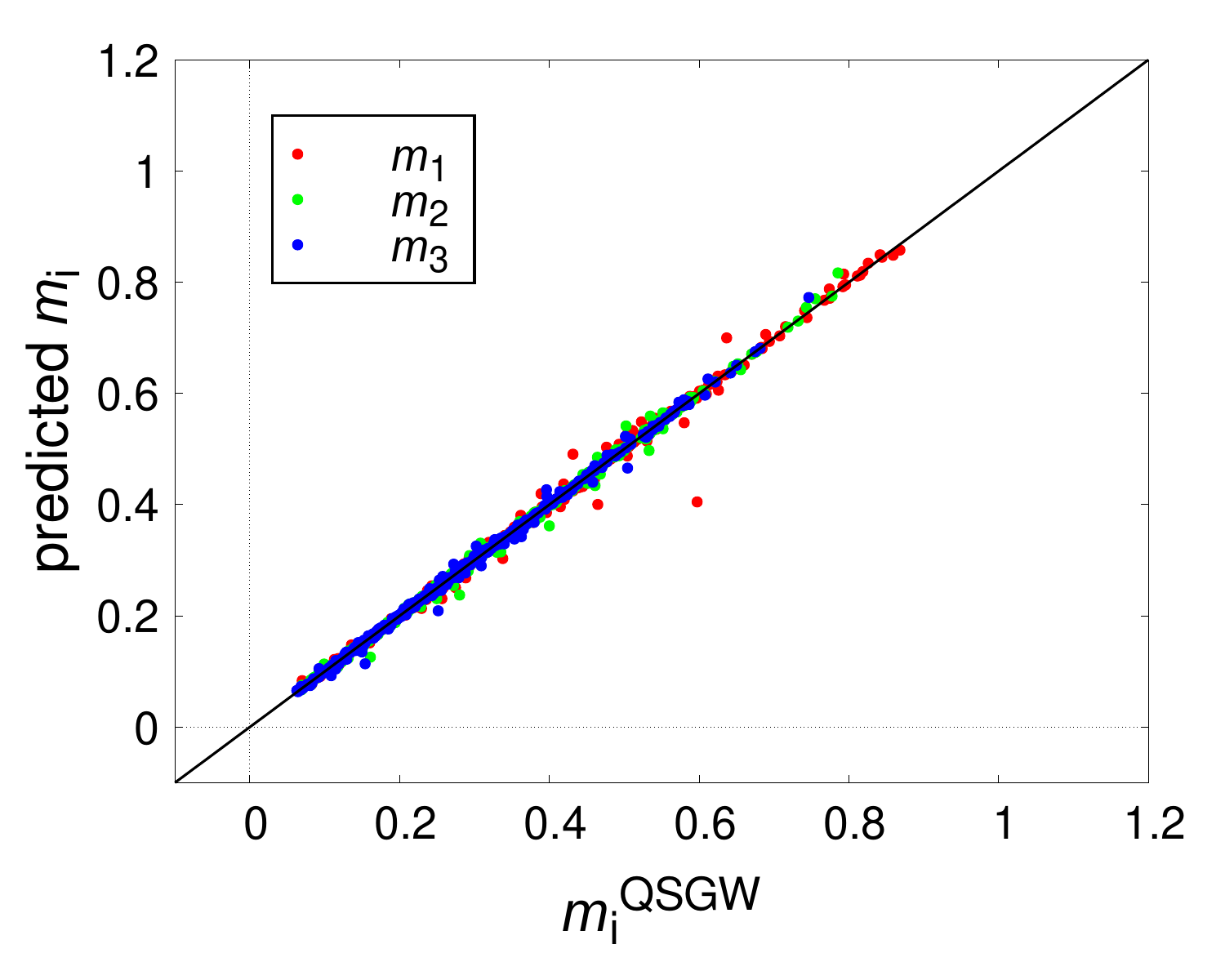}
     \put(-3,73){\textbf{(a)}}
  \end{overpic}
  \hspace{0.5cm}
  \begin{overpic}[width=0.45\columnwidth,clip]{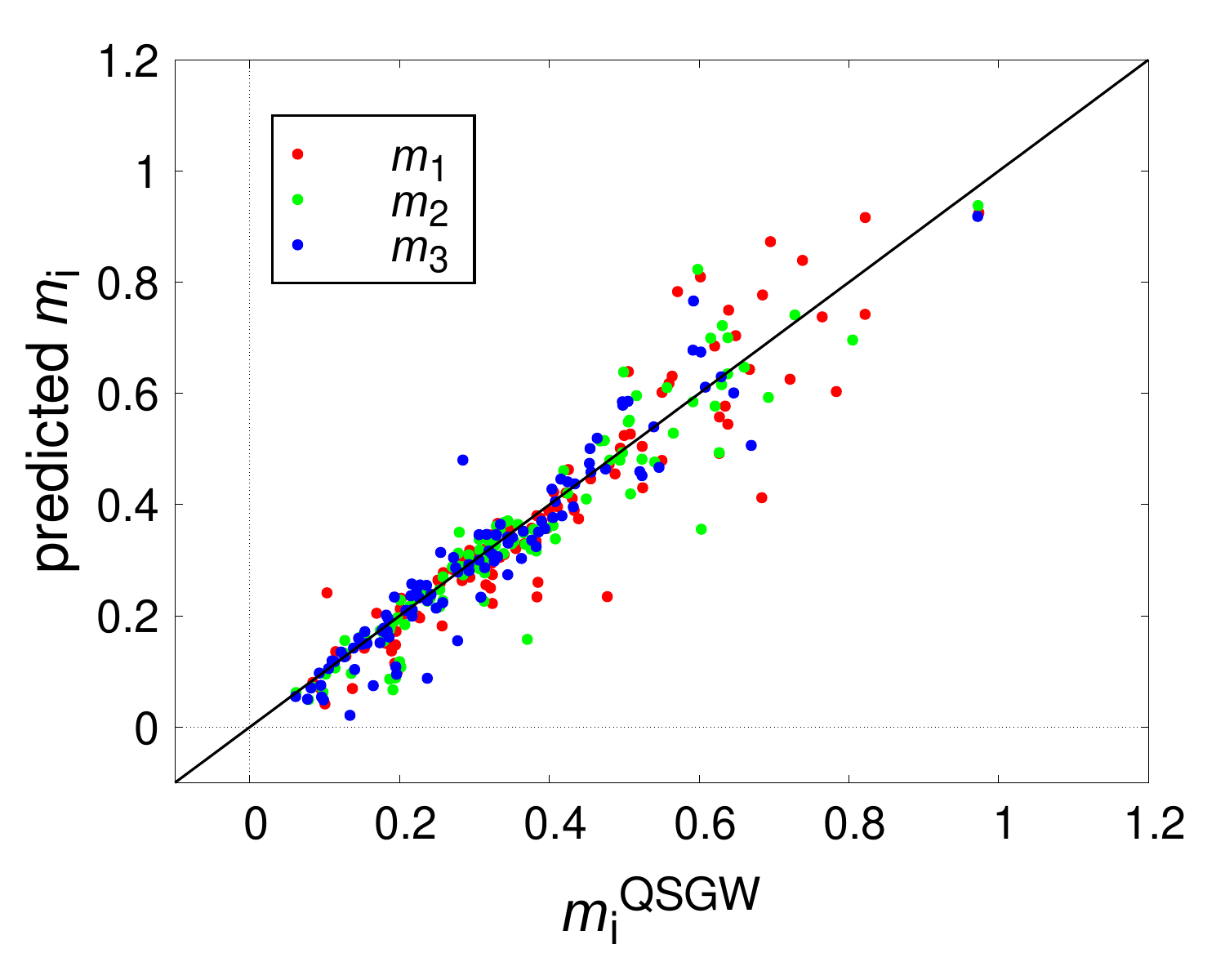}
     \put(-3,73){\textbf{(b)}}
  \end{overpic}
  \caption{
    Results of DOSnp for electron effective mass in units of the electron mass. The inverse of the three eigenvalues 
    of the inverse mass tensor is plotted as $m_1 \geq m_2 \geq m_3$. Panel (a) shows the training results, while 
    panel (b) shows the test results. Predicted $m_3$ in panel (b) is generally on the diagonal line but tends to 
    underestimate the mass in some cases.
  }
  \label{fig:meff_plot}
\end{figure}

For this dataset, we performed training and testing similar to the band gap case. In Fig.~\ref{fig:meff_plot}, we 
show the results for the effective mass ($m_1 \geq m_2 \geq m_3$). Panel (a) shows the training results, while 
panel (b) shows the test results. For the lightest mass $m_3$, we observe relatively narrow scattering along the 
diagonal line in panel (b). However, several predicted $m_3$ values are too small, particularly around 
$m_3^{\rm QSGW} = 0 \sim 0.3$. This indicates that the current model has limitations. To improve the model, 
we need a better approach to handle the three eigenvalues of the mass matrix and to enlarge the dataset.

\begin{table}[H]
   \begin{tabular}{c c c c c c c c c} \hline
\centering
        MPid & $N_a$ & Composition & space group &  & LDA & QSGW & DOSnp & Literatures\\ \hline\hline
        1105229 &4 & ${\rm \beta-GaCuO_2}$  & ${\rm Pna2_1}$         & $E_g$(eV)              & 0.10 & 0.61 & 1.81 & 1.47\cite{GaCuO2} \\ \hline
        1243   &10 & ${\rm \alpha-Ga_2O_3}$ & ${\rm R\overline{3}c}$ & $E_g$(eV)              & 2.73 & 5.53 & 5.38 & 5.3\cite{Ga2O3} \\
             &   &                        &                        & $m_e^*$  & 0.23 & 0.24 & 0.26 & 0.3-0.32\cite{alphaGa2O3}\\ \hline
      886    &10 & ${\rm \beta-Ga_2O_3}$  & C2/m                   & $E_g$(eV)              & 2.16 & 5.22 & 5.07 & 4.9\cite{Ga2O3} \\
             & &                        &                        & $m_e^*$  & 0.23 & 0.25 & 0.28 & 0.28\cite{betaGa2O3}\\ \hline
      1143   &10 & ${\rm \alpha-Al_2O_3}$ & ${\rm R\overline{3}c}$ & $E_g$(eV)              & 5.87 & 9.69 & 9.59 & 8.8\cite{Ga2O3} \\
             &  &                        &                        & $m_e^*$  & 0.38 & 0.35 & 0.37 & 0.3\cite{Al2O3}\\ \hline
      22323  &10 & ${\rm \alpha-In_2O_3}$ & ${\rm R\overline{3}c}$ & $E_g$(eV)              & 1.08 & 3.19 & 3.06 & 3.75\cite{In2O3} \\
             & &                        &                        & $m_e^*$  & 0.15 & 0.19 & 0.19 & 0.18-0.24\cite{In2O3}\\ \hline
      542734 &10 & ${\rm \alpha-Rh_2O_3}$ & ${\rm R\overline{3}c}$ & $E_g$(eV)              & 0.54 & 1.75 & 1.69 & 1.41\cite{Rh2O3}\\ \hline
      11794   &16& ${\rm \beta-AlAgO_2}$  & ${\rm Pna2_1}$         & $E_g$(eV)              & 0.71 & 3.64 & 3.16 & 2.95\cite{AlAgO2} \\ 
              & &                        &                        & $m_e^*$  & 0.30-0.43 & 0.32-0.42 & 0.33-0.41 & - \\ \hline
      1096976 &16& ${\rm \alpha-GaAgO_2}$ & R3m                    & $E_g$(eV)              & 0.25 & 2.12 & 2.06 & 2.4\cite{AlAgO2} \\
              & &                        &                        & $m_e^*$  & 0.17 & 0.19-0.26 & 0.22 & 0.27-0.42\cite{Suzuki_PhD}$^U$ \\ \hline
      1105293 &16 & ${\rm \beta-GaAgO_2}$  & ${\rm Pna2_1}$         & $E_g$(eV)              & 0.21 & 2.53 & 2.10 & 2.1\cite{betaGaAgO2} \\
              & &                       &                        & $m_e^*$  & 0.16-0.27 & 0.22-0.28 & 0.24-0.30 & 0.14-0.33\cite{Suzuki_PhD}$^U$ \\ \hline
   \end{tabular}
\caption{Comparison of LDA and  QSGW for band gap and effective electron mass, together with
the values of our machine learning model DOSnp. We expect DOSnp to reproduce QSGW. It works well overall, however,
we see a large difference for $\beta$-GaCuO$_2$.  
As a reference, we show experimental values taken from the literature together, while the effective mass with superscript $U$ 
is computaional value in LDA+$U$. When effective mass is anisotropic, we show a range of minimum and maximum effective mass.
$N_a$ is the number of atoms in the primitive cell.}
\label{tab:cnn_oxide_expt}
\end{table}

To examine the ability of our DOSnp model further, we applied it to several interesting crystal structures 
for which experimental values are available. In Table~\ref{tab:cnn_oxide_expt}, we list data for crystal 
structures not included in our training dataset. Overall, the values predicted by DOSnp show good agreement 
with those calculated using QSGW. The QSGW values were obtained using the same methodology as when 
generating our dataset. For example, the differences between the two types of $Ga_2O_3$ are well captured 
by QSGW and reproduced by DOSnp, consistent with the known overestimation of band gaps in QSGW.

However, we found an exception: for ${\rm \beta-GaCuO_2}$, QSGW predicts a band gap of 0.61 eV, while 
DOSnp predicts 1.81 eV. This discrepancy remains unexplained. Similar to the case of AlPO$_4$ in 
Fig.~\ref{fig:zoomb}, further investigation is needed to understand the reasons for this difference. Such 
discrepancies between QSGW and DOSnp may provide valuable clues for improving the DOSnp model.

\section{summary}
After applying QSGW to 1,516 crystal structures from the Materials Project (MP), we developed a machine 
learning model, DOSnp, to predict band gaps and effective masses in QSGW. The QSGW calculations were 
performed using the automated scripts \verb#ecalj_auto#, implemented in the \verb#ecalj# package 
\cite{Deguchi2016,Kotani2014,ecalj}. DOSnp is a modification of DOSnet \cite{dosnet}, with the addition of 
a pooling layer that is both logically and practically advantageous compared to the original DOSnet.

We demonstrated that DOSnp outperforms LRLIDG in reproducing QSGW band gaps. Furthermore, DOSnp 
also accurately predicts effective masses. However, we identified cases where DOSnp shows significant 
discrepancies with QSGW. We believe that the current model can be improved by enlarging the dataset and/or 
modifying the model based on these discrepancies. One potential improvement is to include information from 
the crystal orbital Hamilton population (COHP) \cite{COHP}, which provides insights into bonding-antibonding 
characteristics between atoms. Incorporating COHP into DOSnp may enhance its predictive accuracy.

\section{Acknowledgement}
This work was partly supported by JSPS KAKENHI, Grant-in-Aid for Scientific Research (Nos. 22K04909, 24K08229, 23H05457, 23H05449, 24K17608).
We use spglib \cite{spglib} embedded in pymatgen to find crystal symmetries, 
seekpath \cite{seekpath} to find symmetry line for band plots, the material project with pymatgen 
\cite{jain2013,Ong2012b,Ong_2015} are the sources to obtain crystal structures and some basic data sets.
Crystal structure viewer VESTA \cite{Momma2011} was often used on the way on our research.

\section{appendix}
\subsection{ecalj package and automated calculation}
\label{app:ecalj_auto}

We have developed an automated computational system, \verb#ecalj_auto#, for performing QSGW calculations 
on crystal structures from the Materials Project (MP). \verb#ecalj_auto# is a subsystem of the \verb#ecalj# package. 
Using the MP API, we first query crystal structures under specific conditions and obtain them in POSCAR format. 
Next, we refine the POSCAR files to ensure consistency with crystal symmetries using spglib embedded in pymatgen.

For these POSCAR files, all QSGW calculations are performed automatically using default settings optimized 
based on the procedures described in Ref.~\cite{Deguchi2016}. The $k$-mesh for self-energy calculations is set 
to $4\times4\times4$ for Si, with convergence checks performed as detailed in Ref.~\cite{Deguchi2016}. The 
details of \verb#ecalj_auto# will be published elsewhere.

Automated calculations like those provided by \verb#ecalj_auto# are essential for modern materials research. 
While the default settings may lead to issues such as poor convergence in some cases, these problems can 
be addressed through machine learning. As far as we know, \verb#ecalj_auto# combined with \verb#ecalj# is one 
of the most advanced systems for automating GW-related calculations. Given the current state of computational 
materials research, it is impractical to manually check each calculation.

After preparing the default settings, \verb#ecalj_auto# submits Python scripts to job scheduling systems like qsub. 
For example, when dividing 1,500 jobs (crystal structures) into 10 processes, we submit 10 qsub jobs. Each job 
typically uses 32 cores (MPI processes). A single process handling 150 jobs usually completes in a few days. 
Band plots and PDOS plots (not used in this paper) are generated automatically. Symmetry lines for band 
structures are determined using seekpath with spglib.

\subsection{Details of DOSnp model}
\label{app:dosnp}

In Sec.~\ref{tab:DOSnp}, we describe the structure of our DOSnp model. The DOS featurizer, detailed in 
Table~\ref{tab:dosfeatruizer}, was optimized through trial and error to balance simplicity and model size. Unlike 
typical CNNs, our DOS featurizer does not increase the number of channels (filters) at the output. While this 
approach limits optimization, it keeps the model compact. Future improvements could involve increasing the 
filter size or replacing the final output layer with alternatives like softmax. However, such modifications may 
require a larger dataset to avoid overfitting.

\begin{table}[H]
   \centering
   \caption{$N_a$ is the number of atoms in the primitive cell. Add $E_g^{\rm LDA}$ to obtain $E_g^{\rm LDA}$ to the output}
   \begin{tabular}{ccc}  \hline
      Layer & size of output & \\ \hline
      Input & $N_a$ $\times$ (16, 1000) & PDOS in LDA for all atoms $1,...N_a$ \\ 
      DOS Featurizer & $N_a$ $\times$ (16, 12) & Output of DOSFeaturizer\\
      Flattened & $N_a$ $\times$ (192) &  Arrange the output of DOS Featurizer in a row.\\
      Average + Max pooling & (384) & Pooling for all atoms in the primitive cell.\\
      Drop out & (384) & we use drop out ratio p=0.3 \\
      FC+ReLU & (100) & \\
      Linear & (1) & OUTPUT: $E_g^{\rm QSGW} - E_g^{\rm LDA}$ \\ \hline
   \end{tabular}
   \label{tab:DOSnp}
\end{table}
\vspace{-3mm}
\begin{table}[H]
   \centering
      \caption{DOS Featurizer applied to each atom. 1D-CNN is made from a linear transformation followed by RELU activation.}
      \begin{tabular}{lcc}  \hline
         Layer & size of output & (filters, stride, padding) \\ \hline
         Input & ( 16, 1000) & - \\
         1D-CNN & (16, 498) & (5, 2, 0) \\
         Ave. pooling & (16, 249) & (3, 2, 1)  \\
         1D-CNN & (16, 124) & (3, 2, 0)  \\
         Ave. pooling & (16, 62) & (3, 2, 1)  \\
         1D-CNN & (16, 60) & (3, 1, 0)  \\
         Ave. pooling & (16, 30) & (3, 2, 1)  \\
         1D-CNN & (16, 28) & (3, 1, 0)  \\
         Ave. pooling & (16, 14) & (3, 2, 1)  \\
         1D-CNN & (16, 12) & (3, 1, 0)  \\ \hline
     \end{tabular}
     \label{tab:dosfeatruizer}
\end{table}

\subsection{Details of LRLIDG}
\label{app:LRLIDG}
The linear regression (LR) combined with the linearly independent descriptor generation (LIDG) \cite{LIDG} 
enables robust regression by including non-linear terms. We perform LR to minimize the least squares of the 
difference between the predicted and actual band gaps.

As primitive descriptors, we use the following:
- Formation energy
- LDA band gap
- Volume per atom
- Atomic number
- Group of atoms
- Atomic radius
- Atomic mass
- Electronegativity
- First ionization energy

These descriptors are extracted for atoms in the primitive cell. Using LIDG, we generate linearly independent 
descriptors, considering up to second-order products of the primitive descriptors. Except for the LDA band gap, 
the dataset is obtained from the Materials Project (MP) via the API of pymatgen. The LIDG code is available at 
[https://github.com/Hitoshi-FUJII/LIDG](https://github.com/Hitoshi-FUJII/LIDG).

\bibliographystyle{unsrt}
\bibliography{main}
\end{document}